\documentclass[prc,twocolumn,showpacs,floatfix,nofootinbib,%
                superscriptaddress]{revtex4}
\pdfoutput=1   

\usepackage{graphicx}  %
\usepackage{bm}  %
\usepackage{amsmath,amssymb}

\newcommand{\beqn}{\begin{equation}}
\newcommand{\eeqn}{\end{equation}}
\newcommand{\bea}{\begin{eqnarray}}
\newcommand{\eea}{\end{eqnarray}}
\newcommand{\ba}{\begin{align}}
\newcommand{\ea}{\end{align}}

\newcommand{\vlowk}{V_{{\rm low}\,k}}

\newcommand{\fm}{\, \text{fm}}
\newcommand{\fmi}{\, \text{fm}^{-1}}

\newcommand{\Hzero}{H^{\rm bd}}
\newcommand{\flow}{s}

\newcommand{\Trel}{T_{\rm rel}}

\newcommand{\energy}[1]{E_{#1}}
\newcommand{\energyke}[1]{\epsilon_{#1}}

\begin{document}


\title{Block Diagonalization using SRG Flow Equations}

\author{E.\ Anderson}
\affiliation{Department of Physics, The Ohio State University, Columbus, OH 43210}
\author{S.K.\ Bogner}
\affiliation{National Superconducting Cyclotron Laboratory and Department 
  of Physics and Astronomy, Michigan State University, East Lansing, MI 48844}
\author{R.J.\ Furnstahl}
\affiliation{Department of Physics, The Ohio State University, Columbus, OH 43210}
\author{E.D.\ Jurgenson}
\affiliation{Department of Physics, The Ohio State University, Columbus, OH 43210}
\author{R.J.\ Perry}
\affiliation{Department of Physics, The Ohio State University, Columbus, OH 43210}
\author{A.\ Schwenk}
\affiliation{TRIUMF, 4004 Wesbrook Mall, Vancouver, BC, Canada, V6T 2A3}

\date{\today}

\begin{abstract}
%
By choosing appropriate generators for the Similarity Renormalization Group
(SRG) flow equations, different patterns of decoupling in a Hamiltonian can be
achieved. 
Sharp and smooth block-diagonal forms of phase-shift equivalent nucleon-nucleon 
potentials in momentum space are generated as examples and compared to
analogous low-momentum interactions (``$\vlowk$'').
\end{abstract}
\smallskip

\maketitle


The Similarity Renormalization Group
(SRG)~\cite{Glazek:1993rc,Wegner:1994,Kehrein:2006} 
applied to inter-nucleon 
interactions is a continuous series
of unitary transformations implemented as a flow equation for the
evolving Hamiltonian $H_\flow$,
\beqn
   \frac{dH_\flow}{d\flow} = 
    [ \eta_s, H_\flow ]
   = [ [G_s, H_\flow], H_\flow] 
    \,.
   \label{eq:commutator}
\eeqn 
Here $s$ is a flow parameter and
the flow operator $G_s$ specifies the type of SRG~\cite{Bogner:2006srg}. 
Decoupling between low-energy and high-energy matrix elements is naturally 
achieved  in a momentum basis
by choosing a momentum-diagonal flow operator such as the kinetic
energy $\Trel$ or the diagonal of $H_s$; either drives the 
Hamiltonian toward \emph{band-diagonal} form. 
This decoupling leads to dramatically 
improved variational convergence in few-body nuclear systems 
compared to 
unevolved phenomenological or chiral
EFT potentials~\cite{Bogner:2007srg,Bogner:2007rx}.

\begin{figure}[b!]
\includegraphics*[height=3.95cm]{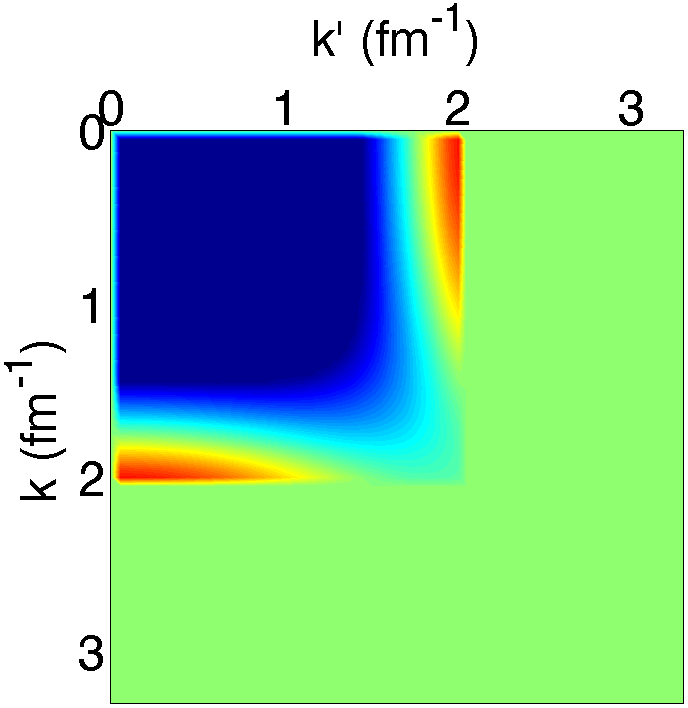}%
~\raisebox{-2pt}{%
\includegraphics*[height=4.05cm]{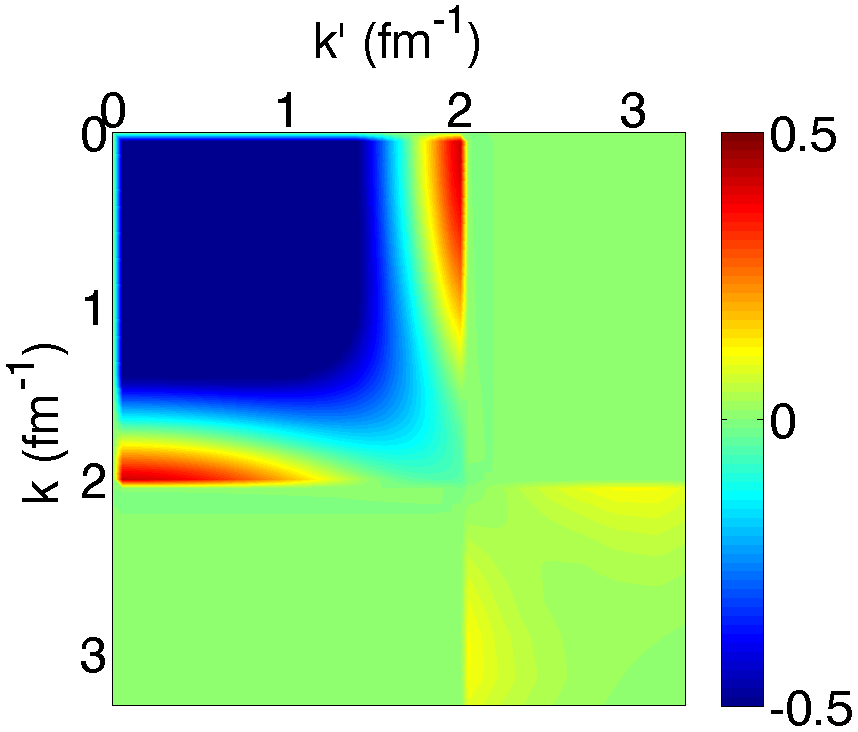}}
\vspace*{-.1in}
\caption{(Color online) Comparison of momentum-space
$\vlowk$ (left) and SRG (right) 
block-diagonal
potentials with $\Lambda = 2\fmi$ 
evolved from an N$^3$LO $^3$S$_1$ potential~\cite{N3LO}. The color axis
is in fm.}
\label{fig:vlowk}
\end{figure}

\begin{figure}[b!]
\includegraphics*[width=3.75cm]{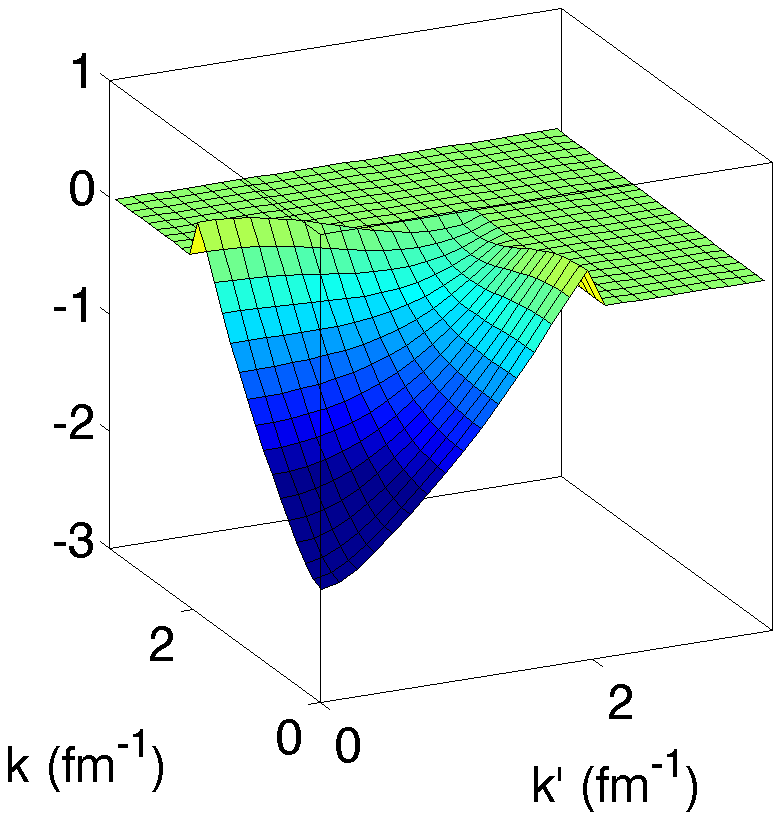}%
\raisebox{0pt}{%
\includegraphics*[width=4.85cm]{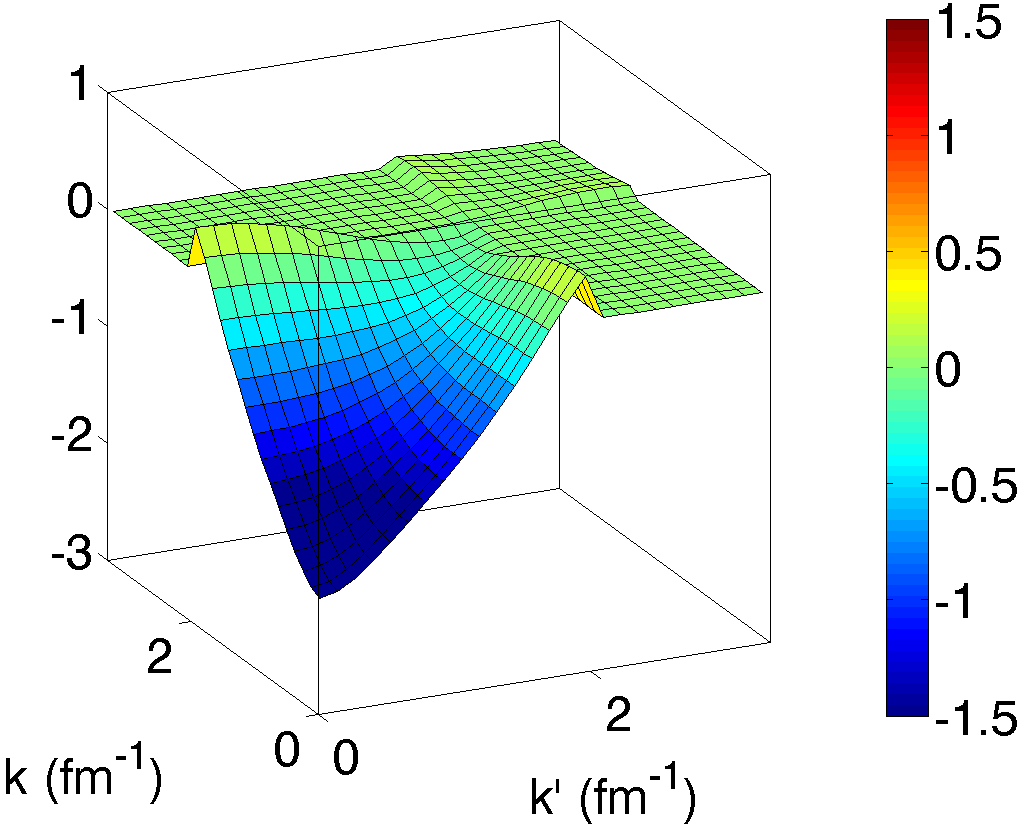}}
\vspace*{-.1in}
\caption{(Color online) Comparison of momentum-space
$\vlowk$ (left) and SRG (right) 
block-diagonal
potentials with $\Lambda = 2\fmi$
 evolved from an N$^3$LO $^3$S$_1$ potential~\cite{N3LO}. The color and
 $z$ axes
are in fm.}
\label{fig:vlowksurface}
\end{figure}

Renormalization Group (RG) methods that evolve NN interactions with
a sharp or smooth cutoff in relative momentum,
known generically as $\vlowk$,
rely on the invariance of the two-nucleon T matrix~\cite{Vlowk2,Bogner:2006vp}.
These approaches achieve a \emph{block-diagonal} form characterized by a
cutoff $\Lambda$ (see left plots in
Figs.~\ref{fig:vlowk} and \ref{fig:vlowksurface}).
As usually 
implemented they set the high-momentum matrix elements to zero
but this is not required.

\begin{figure*}[tb!]
\includegraphics*[width=4cm]{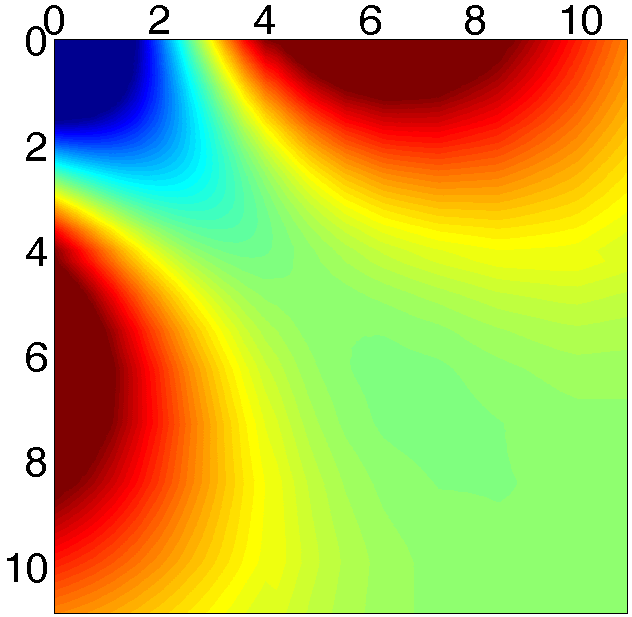}
\includegraphics*[width=4cm]{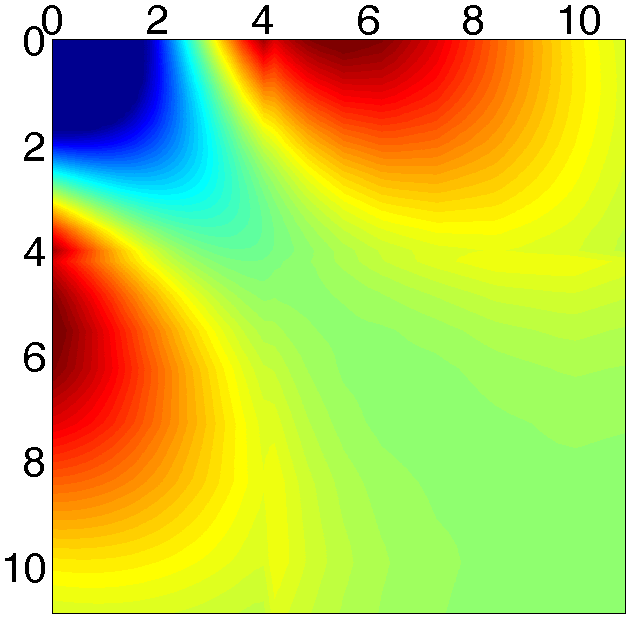}
\includegraphics*[width=4cm]{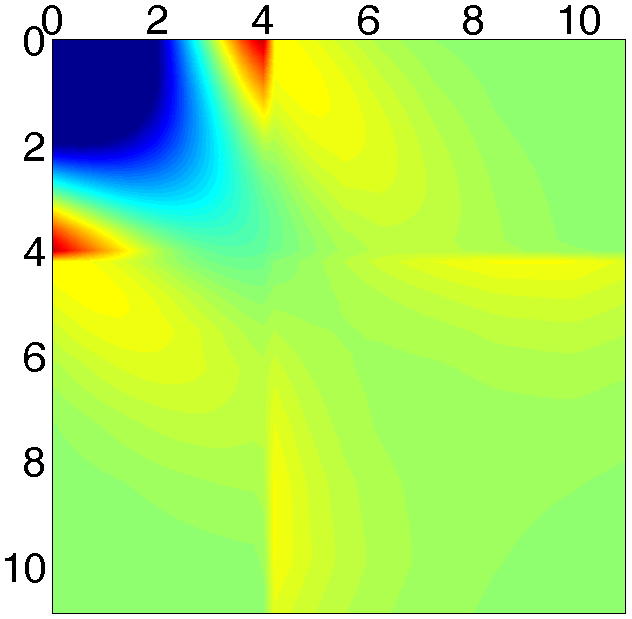}
\includegraphics*[width=4cm]{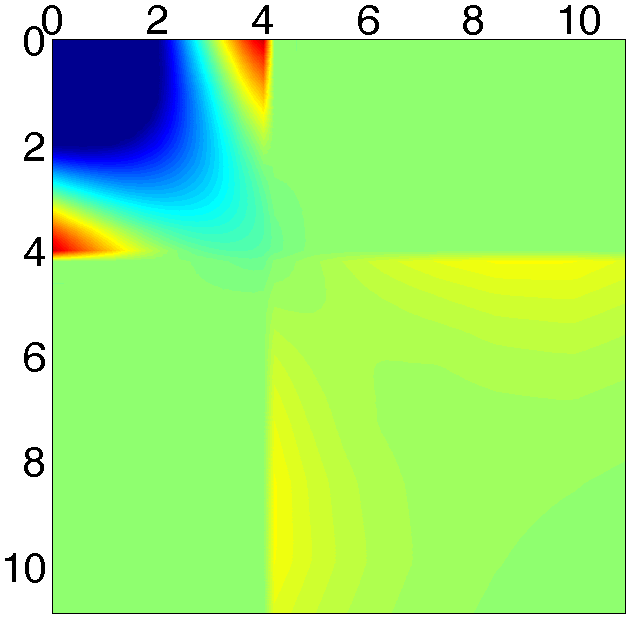}
\vspace*{-.1in}
\caption{(Color online) Evolution of the $^3$S$_1$ partial 
wave with a sharp block-diagonal flow equation with $\Lambda =2\fmi$
at $\lambda = 4$, 3, 2, and $1\fmi$. 
The initial N$^3$LO potential is from Ref.~\cite{N3LO}.
The axes are in 
units of $k^2$ from 0--11 fm$^{-2}$. 
The color scale ranges from  $-0.5$ to $+0.5\fm$ as in Fig.~\ref{fig:vlowk}.}
\label{fig:bd_srg_sharp_3s1}
\end{figure*}

\begin{figure*}[htb!]
\includegraphics*[width=4cm]{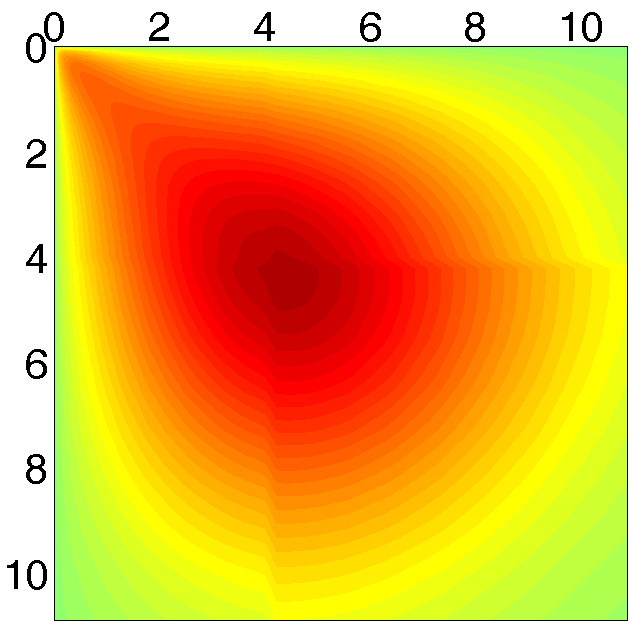}
\includegraphics*[width=4cm]{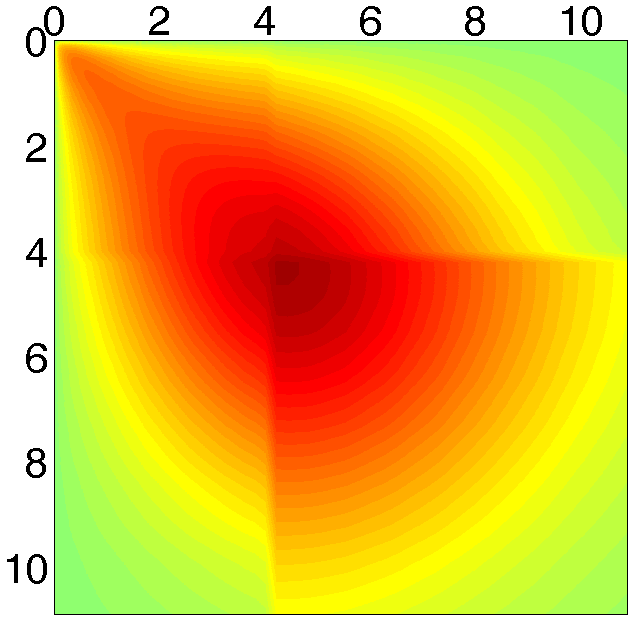}
\includegraphics*[width=4cm]{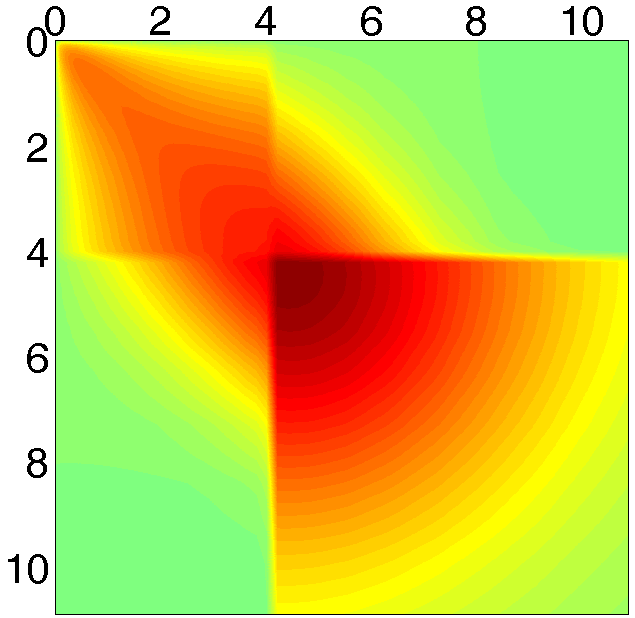}
\includegraphics*[width=4cm]{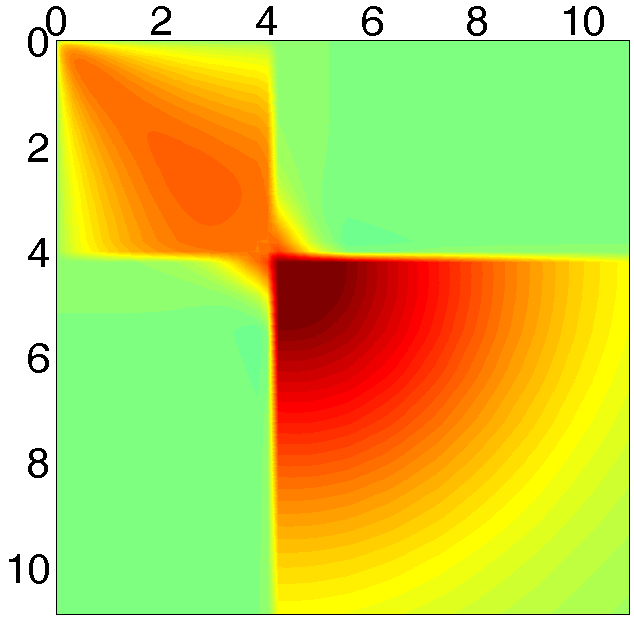}
\vspace*{-.1in}
\caption{(Color online) Same as Fig.~\ref{fig:bd_srg_sharp_3s1} but
for the $^1$P$_1$ partial 
wave.}
\label{fig:bd_srg_sharp_1p1}
\end{figure*}

Block-diagonal decoupling
of the sharp $\vlowk$ form can be generated using SRG 
flow equations by choosing a block-diagonal flow 
operator~\cite{Elena99,Gubankova:2000cia}, 
\beqn
  G_s =  \left( 
        \begin{array}{cc}
          PH_{s}P & 0   \\
          0     & QH_{s}Q
        \end{array}
        \right) 
       \equiv \Hzero_s 
        \;,
     \label{eq:Hbd}   
\eeqn
with projection operators $P$ and $Q = 1 - P$. 
In a partial-wave momentum
representation, $P$ and $Q$ are step functions defined by a sharp cutoff
$\Lambda$ on relative momenta.
This choice for $G_s$, which means that $\eta_s$ is non-zero only where
$G_s$ is zero,
suppresses off-diagonal matrix elements such that the
Hamiltonian approaches a block-diagonal form as $s$ increases.
If one considers a measure of the off-diagonal coupling of the Hamiltonian,
\beqn
  {\rm Tr}[(Q H_s P)^{\dagger} (Q H_s P)]
    = {\rm Tr}[P H_s Q H_s P] \geqslant 0
    \;,
    \label{eq:QHPmeasure}
\eeqn
then its derivative is easily evaluated by applying the SRG 
equation, Eq.~(\ref{eq:commutator}):
\bea
  &&  \frac{d}{ds} {\rm Tr}[P H_s Q H_s P]
   \nonumber \\
    && \qquad = 
    {\rm Tr}[P\eta_s  Q(Q H_s Q H_s P - Q H_s P H_s P)]
    \nonumber \\
    & & \qquad\qquad\null +
        {\rm Tr}[(P H_s P H_s Q - P H_s Q H_s Q) Q\eta_s  P]
   \nonumber \\
   && \qquad 
     = -2 {\rm Tr} [(Q \eta_s P)^{\dagger} (Q \eta_s  P)]
     \leqslant 0
     \;.
\eea
Thus, the off-diagonal $Q H_s P$ block will decrease in general
as $s$ increases~\cite{Elena99,Gubankova:2000cia}.

The right plots in Figs.~\ref{fig:vlowk} and \ref{fig:vlowksurface}
result from evolving the N$^3$LO potential from
Ref.~\cite{N3LO} using the block-diagonal $G_s$ of Eq.~(\ref{eq:Hbd})
with $\Lambda = 2\fmi$ until $\lambda \equiv 1/s^{1/4} = 0.5\fmi$.
The agreement between $\vlowk$ and SRG potentials for momenta
below $\Lambda$ is striking.
A similar degree of universality is found in the other partial waves.
Deriving an explicit connection between these approaches is the
topic of an ongoing investigation.

The evolution with $\lambda$ of two representative partial waves 
($^3$S$_1$ and $^1$P$_1$) 
are shown in
Figs.~\ref{fig:bd_srg_sharp_3s1} and \ref{fig:bd_srg_sharp_1p1}.   
The evolution of the ``off-diagonal'' 
matrix elements (meaning those outside the $PH_sP$ and $QH_sQ$ blocks)  
can be roughly understood from the
dominance of the kinetic energy on the diagonal.
Let the indices $p$ and $q$ run over indices of the momentum
states in the $P$ and $Q$ spaces, respectively.
To good approximation we can replace $P H_s P$ and
$Q H_s Q$ by their eigenvalues $E_p$ and $E_q$ 
in the SRG equations, yielding~\cite{Elena99,Gubankova:2000cia}
\beqn
  \frac{d}{ds}h_{pq} \approx \eta_{pq} \energy{q} - \energy{p}\eta_{pq}
    = -(\energy{p}-\energy{q})\, \eta_{pq}
    \label{eq:hpq} 
\eeqn
and   
\beqn
  \eta_{pq} \approx \energy{p} h_{pq} - h_{pq} \energy{q} 
    = (\energy{p} - \energy{q})\, h_{pq} \;.
   \label{eq:etapq}
\eeqn
Combining these two results, we have the evolution of any
off-diagonal matrix element:
\beqn
  \frac{d}{ds}h_{pq} \approx - (\energy{p} - \energy{q})^2 \, h_{pq}
  \;. \label{eq:hpqapprox}
\eeqn
In the NN case we can replace the eigenvalues by those for the relative 
kinetic energy,
giving an explicit solution
\beqn
   h_{pq}(s) \approx h_{pq}(0)\, e^{-s(\energyke{p} - \energyke{q})^2}
   \label{eq:explicit}
\eeqn
with $\energyke{p} \equiv p^2/M$.
Thus the off-diagonal elements
go to zero with the
energy differences just like
with the SRG with $T_{\rm rel}$;
one can see the width of order $1/\sqrt{s} = \lambda^2$ in the $k^2$ plots
of the evolving potential
in Figs.~\ref{fig:bd_srg_sharp_3s1} and \ref{fig:bd_srg_sharp_1p1}.

\begin{figure*}[htb!]
\includegraphics*[width=4cm]{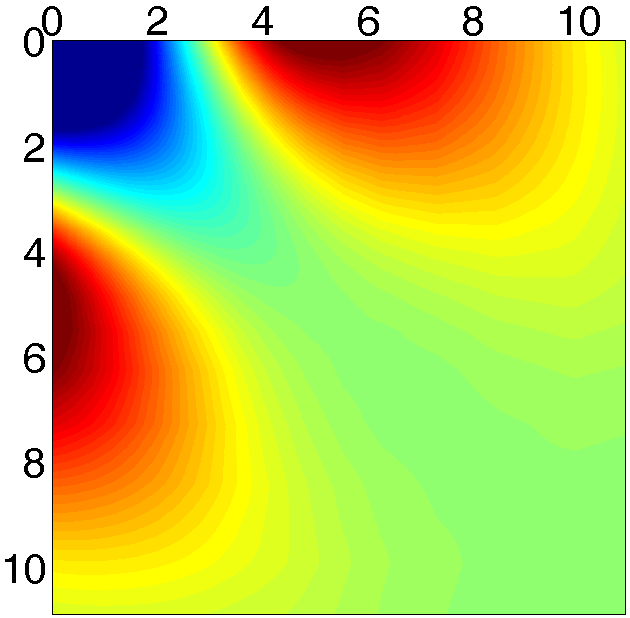}
\includegraphics*[width=4cm]{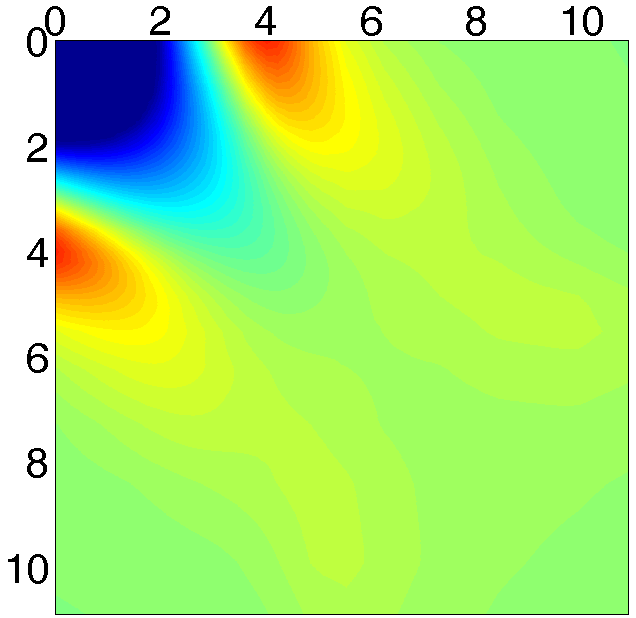}
\includegraphics*[width=4cm]{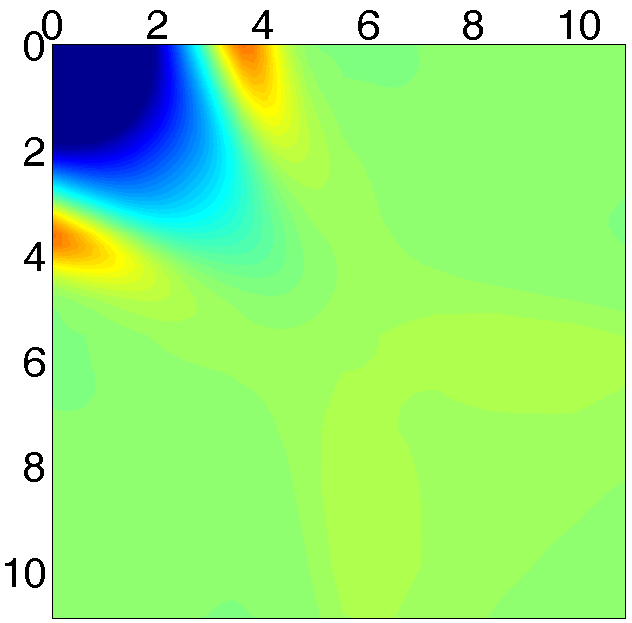}
\includegraphics*[width=4cm]{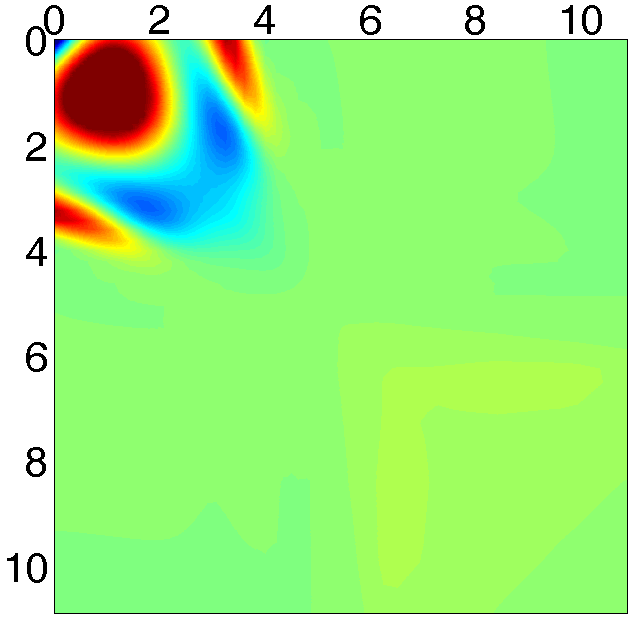}
\vspace*{-.1in}
\caption{(Color online) Evolution of the $^3$S$_1$
partial  wave with a smooth ($n=4$) block-diagonal flow equation with
$\Lambda =2.0\fmi$, starting with the
N$^3$LO potential from Ref.~\cite{N3LO}. 
The flow parameter $\lambda$ is 
3, 2, 1.5, and $1\fmi$. The  axes are in units of $k^2$ from 0--11
fm$^{-2}$. The color scale ranges from $-0.5$ to $+0.5\fm$ 
as in Fig.~\ref{fig:vlowk}.}
\label{fig:bd_srg_smooth}
\end{figure*}
 
While in principle the evolution to a sharp block-diagonal form
means going to $s = \infty$ ($\lambda = 0$),
in practice
we need only take $s$  as large as needed  to quantitatively achieve the
decoupling implied by Eq.~(\ref{eq:explicit}).
Furthermore, it should hold for more general definitions of $P$ and $Q$.
To smooth out the cutoff, we can introduce a smooth regulator $f_\Lambda$,
which we take here to be an exponential form:
\beqn
   f_\Lambda(k) = e^{-(k^2/\Lambda^2)^n}  \;,
   \label{eq:fLambda}
\eeqn 
with $n$ an integer. For $\vlowk$ potentials, 
typical values used are $n=4$ and $n=8$
(the latter is considerably sharper but still numerically robust).
By replacing $\Hzero_s$ with
\beqn
  G_s = f_\Lambda H_s f_\Lambda + (1-f_\Lambda)H_s(1-f_\Lambda)  \;,
  \label{eq:Gs}
\eeqn
we get a smooth block-diagonal potential.

A representative example with $\Lambda=2\fmi$ and $n=4$ is 
shown in Fig.~\ref{fig:bd_srg_smooth}. 
We can evolve to $\lambda = 1.5\fmi$ without a problem.
For smaller $\lambda$ the overlap of the $P$ and $Q$ spaces
becomes significant and the potential becomes distorted.  
This distortion indicates that there is no
further benefit to evolving in $\lambda$ very far below $\Lambda$;
in fact the decoupling worsens for $\lambda < \Lambda$ with a smooth
regulator.

\begin{figure}[b!]
\includegraphics*[width=6.6cm]{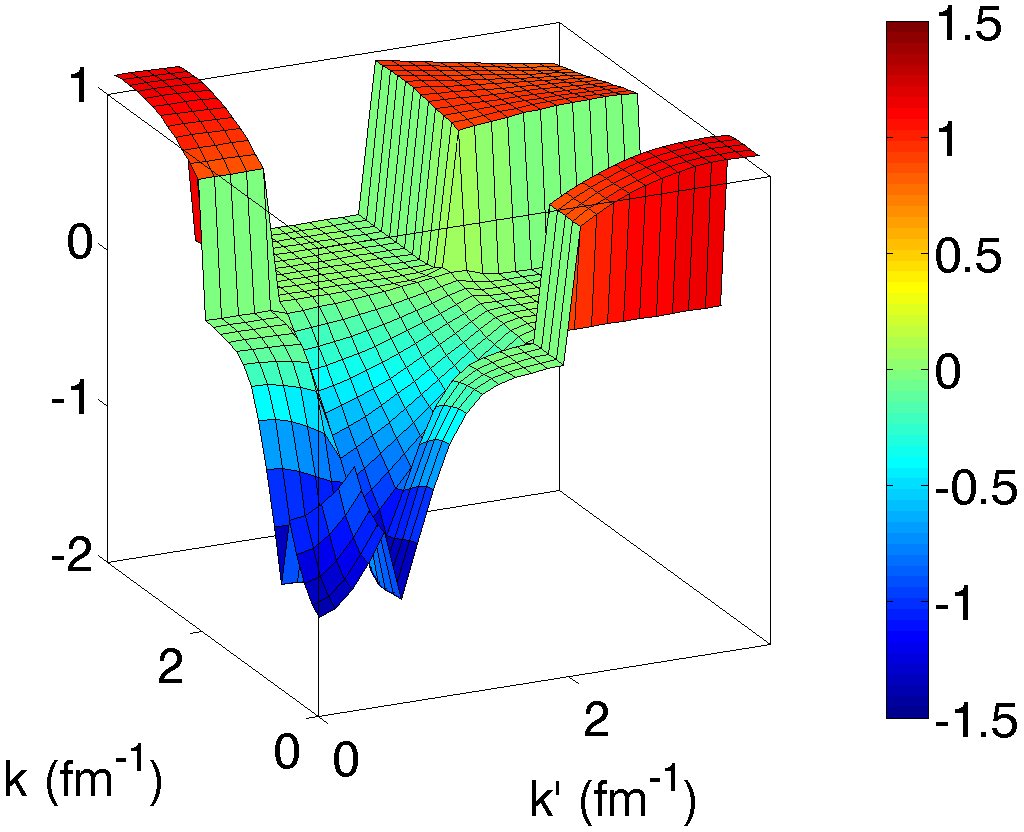}
\includegraphics*[width=6.6cm]{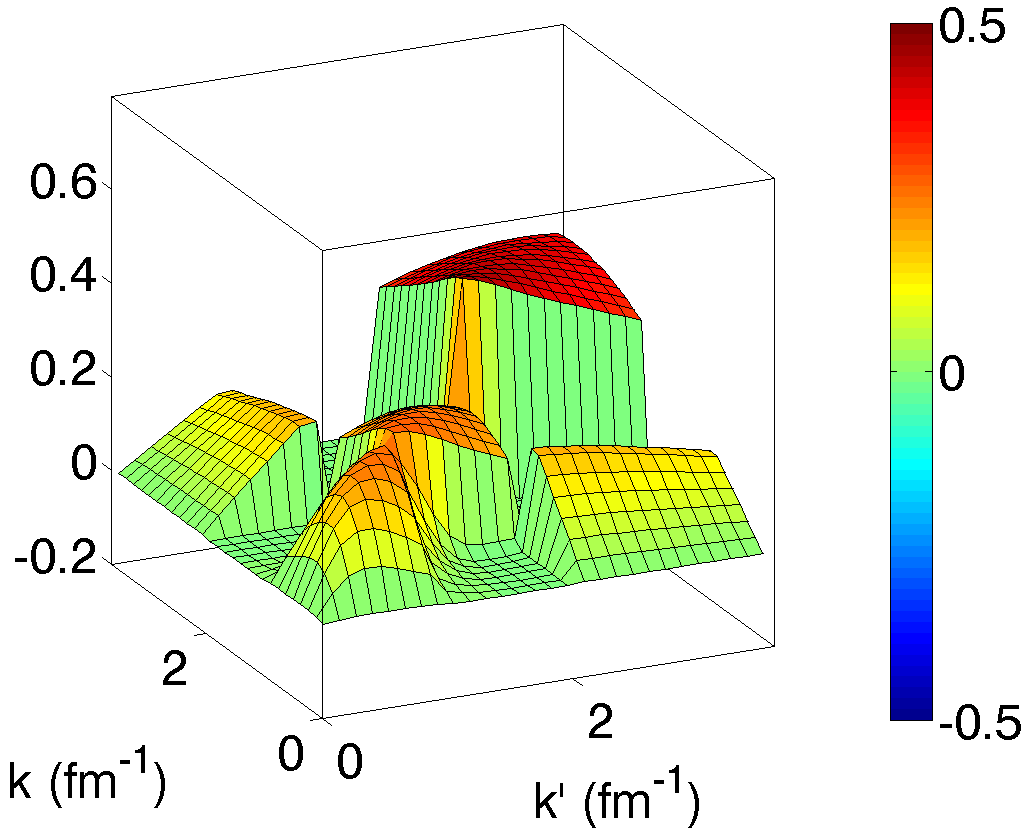}
\vspace*{-.1in}
\caption{(Color online) Evolved SRG potentials starting from
Argonne $v_{18}$ in the $^1$S$_0$ and
$^1$P$_1$ partial waves to $\lambda = 1\fmi$
using a bizarre choice
for $G_s$ (see text). The color and
 $z$ axes
are in fm. }
\label{fig:vsrgweird}
\end{figure}

Another type of SRG that is second-order exact and yields similar
block diagonalization is defined by
\beqn
  \eta_s = [T, PV_sQ + QV_sP] \;,  
\eeqn
which can be implemented with $P \rightarrow f_\Lambda$ and
$Q \rightarrow (1-f_\Lambda)$, with $f_\Lambda$ either sharp
or smooth. 
We can also consider bizarre choices for $f_\Lambda$ in Eq.~(\ref{eq:Gs}), such
as defining it to be zero out to $\Lambda_{\rm lower}$, then unity out to 
$\Lambda$, and then zero above that.  This means that $1 - f_\Lambda$ 
defines both low and high-momentum blocks and the region that 
is driven to zero consists of several rectangles. 
Results for two partial waves starting from the Argonne $v_{18}$
potential~\cite{Wiringa:1994wb} are shown in Fig.~\ref{fig:vsrgweird}.
Despite the strange appearence, these remain unitary transformations of
the original potential,
with phase shifts and other NN observables the same as with the original
potential. 
These choices provide a 
proof-of-principle that the decoupled regions can be tailored 
to the physics problem at hand.


\begin{figure}[bth!]
\includegraphics*[width=6.8cm]{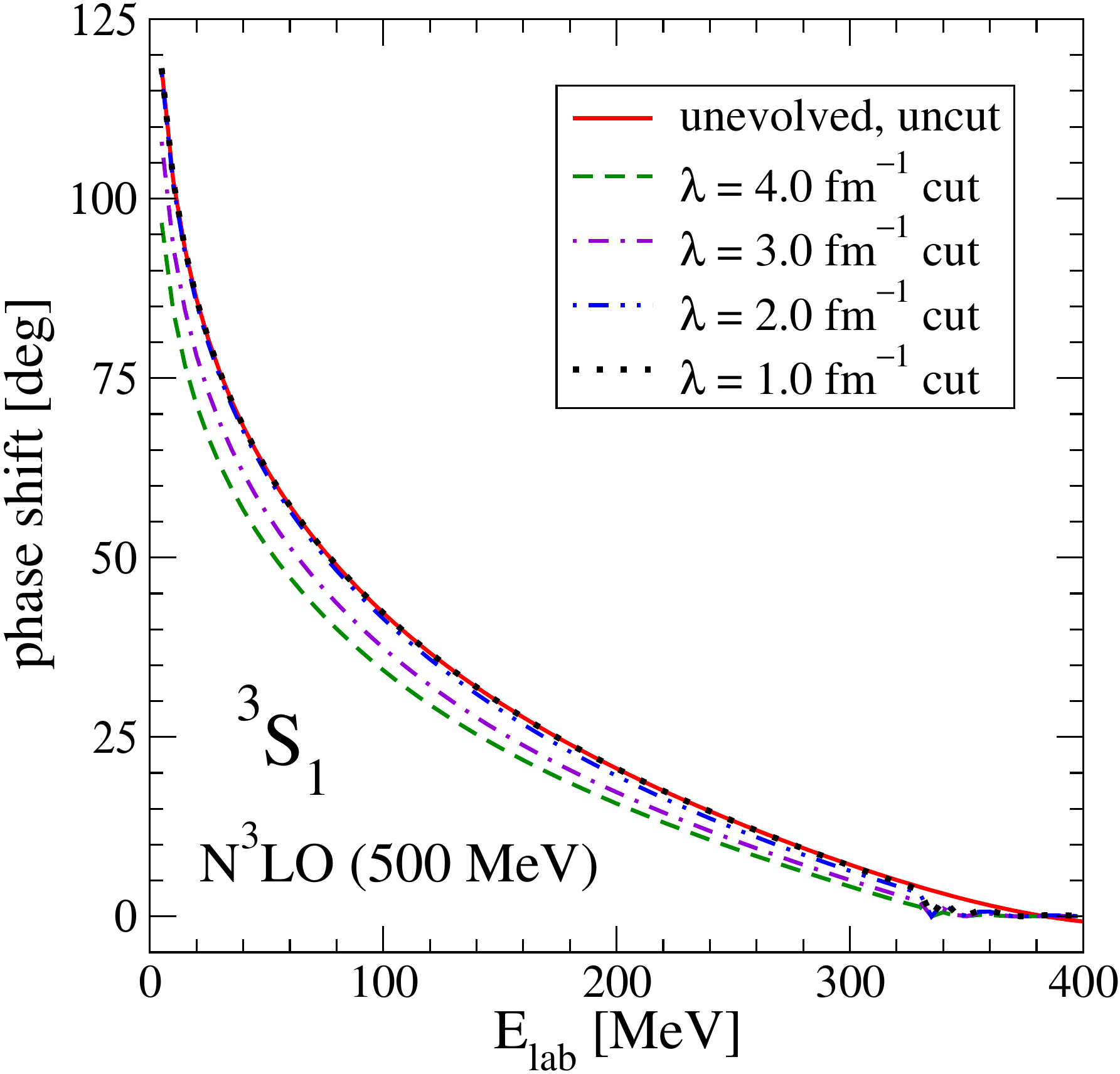}
\vspace*{-.1in}
\caption{(Color online) Phase shifts for the $^3$S$_1$ partial wave
from an initial N$^3$LO potential and 
the evolved sharp SRG block-diagonal potential with $\Lambda=2\fmi$
at various $\lambda$, in each case with the potential set identically
to zero above $\Lambda$.}
\label{fig:phase_sharp}
\end{figure}

\begin{figure}[bth!]
\includegraphics*[width=6.8cm]{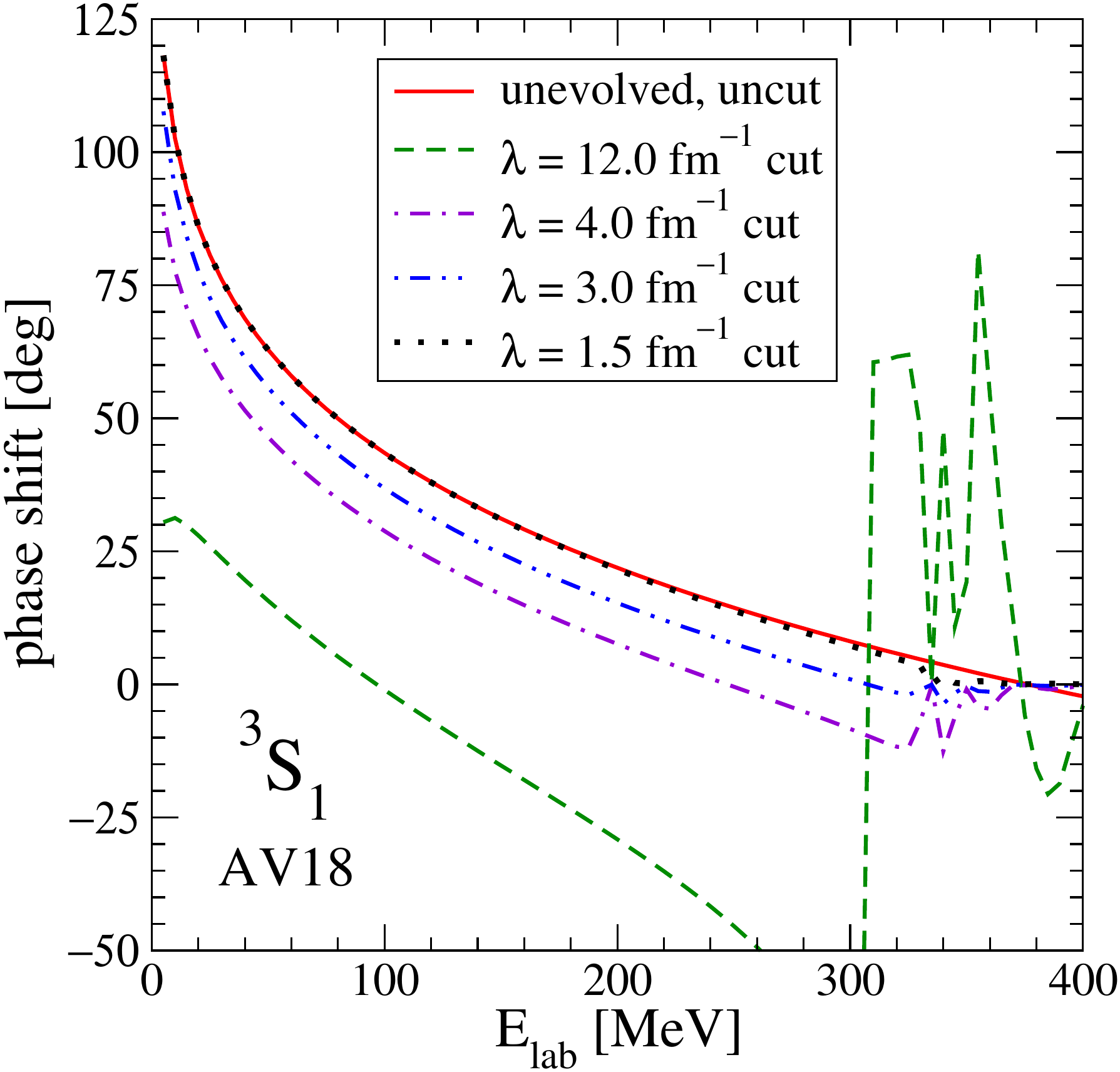}
\vspace*{-.1in}
\caption{(Color online) 
Same as Fig.~\ref{fig:phase_sharp} but with Argonne
$v_{18}$ as the initial potential~\cite{Wiringa:1994wb}.}
\label{fig:phase_sharpAV18}
\end{figure}

\begin{figure}[bth!]
 \includegraphics*[width=6.8cm]{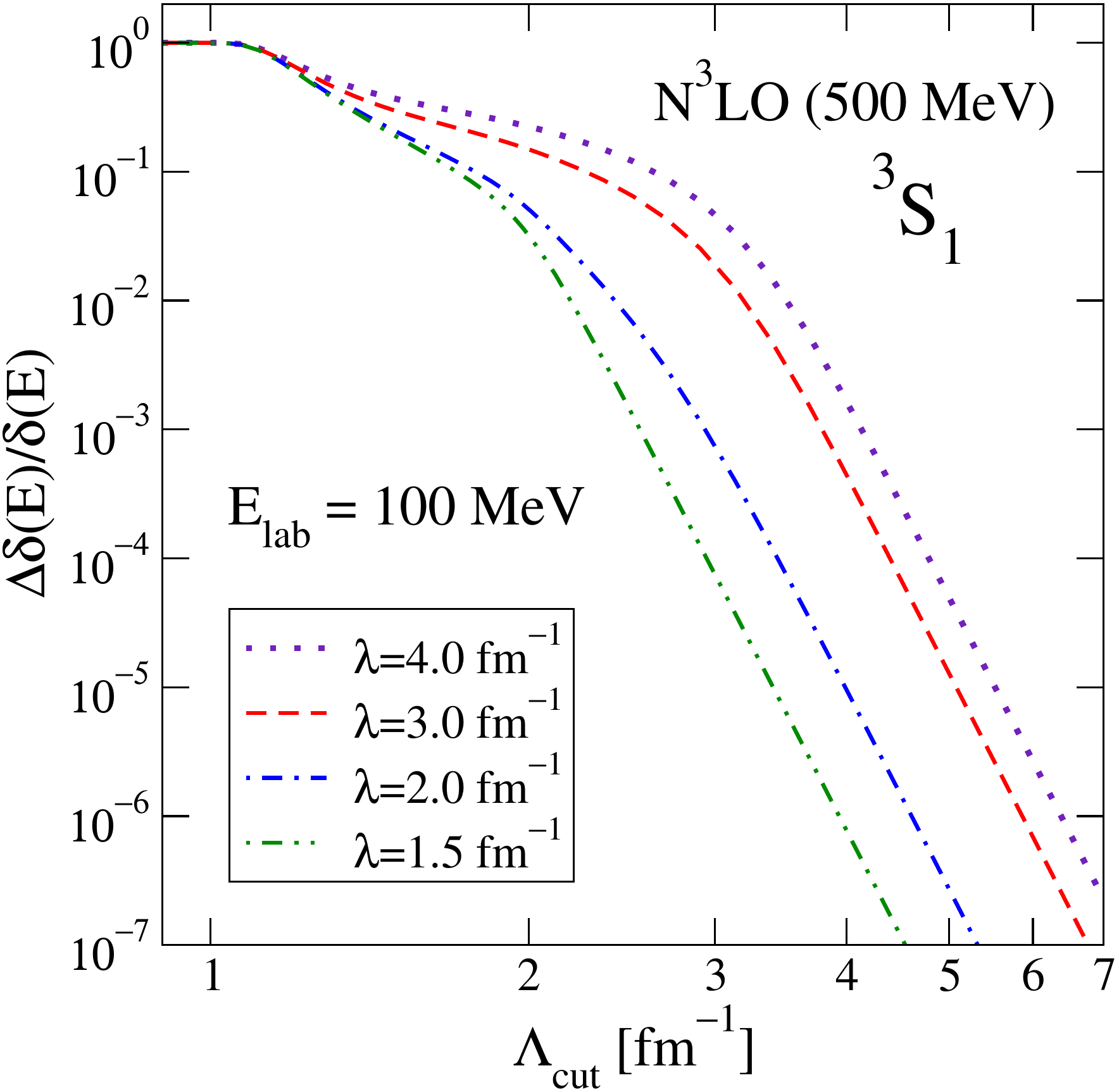}
 \vspace*{-.1in}
 \caption{(Color online) Errors in the phase shift at $E_{\rm lab} = 100\,$MeV
 for the evolved sharp SRG block-diagonal 
 potential with $\Lambda=2\fmi$ for a range
 of $\lambda$'s and a regulator with $n=8$.
 \label{fig:phase_err_sharp}}
\end{figure}

Definitive tests of decoupling for NN observables are now possible for
$\vlowk$ potentials since  the unitary transformation of the SRG
guarantees that no physics is lost.  For example,
in Figs.~\ref{fig:phase_sharp} and \ref{fig:phase_sharpAV18} we
show $^3$S$_1$ phase shifts from an SRG sharp block
diagonalization with $\Lambda = 2\fmi$ for two different potentials. 
The phase shifts  are
calculated with the potentials cut sharply at $\Lambda$. That is, the
matrix  elements of the potential are set to zero above that point. The
improved  decoupling as $\lambda$ decreases is evident in each case. 
By $\lambda = 1\fmi$ in  Fig.~\ref{fig:phase_sharp}, 
the unevolved and evolved curves
are indistinguishable  to the width of the line up to about 300\,MeV. 

In Fig.~\ref{fig:phase_err_sharp} we show a quantitative analysis of the 
decoupling as in Ref.~\cite{decoupling_paper}. The figure shows the relative 
error of the phase shift at 100\,MeV
calculated with a potential that is cut off by a 
smooth regulator as in Eq.~(\ref{eq:fLambda}) 
at a series of values $\Lambda_{\rm cut}$. 
We observe the same universal decoupling behavior seen in 
Ref.~\cite{decoupling_paper}:
a shoulder indicating the perturbative 
decoupling region, where the slope matches the power $2n$ fixed by
the smooth regulator.
The onset of the shoulder in $\Lambda_{\rm cut}$ decreases with $\lambda$
until it saturates 
for $\lambda$ somewhat below $\Lambda$,
leaving the shoulder at $\Lambda_{\rm cut} \approx \Lambda$.
Thus, as $\lambda\rightarrow 0$ the decoupling scale is set
by the cutoff $\Lambda$.

In the more conventional SRG, where
we use $\eta_s = [T,H_s] = [T,V_s]$, it is easy to see that the 
evolution of the two-body potential in the two-particle system can be 
carried over directly to the three-particle system. 
In particular, it follows that the three-body 
potential does not depend on disconnected two-body
parts~\cite{Bogner:2006srg,Bogner:2007qb}. 
If we could implement 
$\eta_s$ as proposed here with analogous properties,
we would have a tractable method for generating $\vlowk$ three-body forces.
While it seems possible to define Fock-space operators 
with projectors $P$ and $Q$ 
that will not have problems with disconnected parts,
it is not yet clear whether 
full decoupling in the few-body space can be realized. 
Work on this problem is in progress.

\vspace*{.1in}

\begin{acknowledgments}
This work was supported in part by the National Science 
Foundation under Grant Nos.~PHY--0354916 and PHY--0653312,  
the UNEDF SciDAC Collaboration under DOE Grant 
DE-FC02-07ER41457,
and the Natural Sciences and Engineering Research Council of Canada 
(NSERC). TRIUMF receives federal funding via a contribution agreement 
through the National Research Council of Canada.
\end{acknowledgments}


\end{document}